\documentclass[aps,prd,preprint,showpacs,eqsecnum,nofootinbib,superscriptaddress]{revtex4}

\usepackage{amssymb,amsmath,psfrag,epsfig}

\def\bra#1#2{[#1\,#2]}
\def\ket#1#2{\langle #1\,#2\rangle}
\def\fig#1{Fig.~{\ref{#1}}}

\begin{document}

\noindent \hfill Brown-HET-1572

\noindent \hfill LAPTH-1291/08

\vskip 1 cm

\title{Tree-Level Amplitudes in ${\mathcal{N}}=8$ Supergravity}

\author{J.~M.~Drummond}

\affiliation{LAPTH, Universit\'e de Savoie, CNRS, Annecy-le-Vieux Cedex,
France}

\author{M.~Spradlin}

\affiliation{Brown University, Providence, Rhode Island 02912, USA}

\author{A.~Volovich}

\affiliation{Brown University, Providence, Rhode Island 02912, USA}

\author{C.~Wen}

\affiliation{Brown University, Providence, Rhode Island 02912, USA}

\begin{abstract}
We present an algorithm for writing down
explicit formulas for all tree amplitudes in
${\mathcal{N}}=8$ supergravity,
obtained from solving
the supersymmetric on-shell recursion relations.
The formula is patterned after one recently obtained for
all tree amplitudes in ${\mathcal{N}}=4$ super Yang-Mills which involves
nested sums of dual superconformal invariants.
We find that all graviton amplitudes can be written in terms of
exactly the same structure of nested
sums with two modifications: the dual superconformal invariants
are promoted from ${\mathcal{N}} = 4$ to ${\mathcal{N}} = 8$ superspace
in the simplest manner possible--by squaring them--and
certain additional non-dual conformal gravity dressing
factors (independent of the superspace coordinates) are
inserted into the nested sums.
To illustrate the procedure we give explicit closed-form formulas
for all NMHV, NNMHV and NNNMV gravity super-amplitudes.
\end{abstract}


\maketitle

\section{Introduction}

The past several years have witnessed dramatic progress in our understanding
of gluon scattering amplitudes, especially in the
maximally supersymmetric ${\mathcal{N}} = 4$ super-Yang-Mills theory (SYM).
These advances have provided a pleasing mix of theoretical insights,
shedding light on the mathematical structure of amplitudes and their
role in gauge/string duality, and more practical results,
including impressive
new technology for carrying
out previously impossible calculations at tree level
and beyond.

It has recently been pointed out~\cite{ArkaniHamed:2008gz} that
there are reasons to suspect ${\mathcal{N}} = 8$ supergravity (SUGRA) to have
even richer structure and to be ultimately even simpler than
SYM. 
Despite great
progress~\cite{Bern:1993wt,Bern:1998ug,Bern:2002kj,Giombi:2004ix,Nair:2005iv,Cachazo:2005ca,BjerrumBohr:2005jr,BjerrumBohr:2006sg,Bern:2006kd,Benincasa:2007qj,Bern:2007hh,ArkaniHamed:2008yf,Kallosh:2008ic,Naculich:2008ys,Bianchi:2008pu,Naculich:2008ew,BjerrumBohr:2008ji,Bern:2008qj,Kallosh:2008mq,Bern:2008pv,Mason:2008ms,Badger:2008rn,Kallosh:2008rr,Kallosh:2008ru,Spradlin:2008bu}
however,
our understanding of SUGRA amplitudes is still poor compared to SYM,
suggesting that we are still missing some key insights
into this problem.

Nowhere is the disparity between
our understanding of SYM and SUGRA more transparent
than in the expressions for what should be their simplest nontrivial scattering
amplitudes, those describing the interaction of 2 particles of one
helicity with $n-2$ particles of the opposite helicity.
In SYM these maximally helicity violating (MHV) amplitudes are
encapsulated in the stunningly simple formula conjectured by
Parke and Taylor~\cite{Parke:1986gb} and proven by Berends and
Giele~\cite{Berends:1987me}, which we express here (as throughout
this paper) in on-shell ${\mathcal{N}} = 4$
superspace
\begin{equation}
\label{AMHV}
A^{\rm MHV}(1,\ldots,n) = \frac{ \delta^{(8)}(q) }{ \ket{1}{2} \ket{2}{3}
\cdots \ket{n}{1} }\,.
\end{equation}
In contrast, all known explicit formulas for $n$-graviton MHV amplitudes are
noticeably more complicated.  The first such formula was conjectured
20 years ago~\cite{Berends:1988zp} and a handful of alternative
expressions of more or less the same degree of complexity have appeared
more recently~\cite{Bedford:2005yy,Bern:2007xj,Elvang:2007sg,Spradlin:2008bu}.

Beyond MHV amplitudes the situation is even less satisfactory,
though the Kawai-Lewellen-Tye (KLT)
relations~\cite{Kawai:1985xq}
may be used in principle to express any desired amplitude as a complicated
sum of various permuted squares of gauge theory amplitudes and other factors.
These relations are a consequence of the relation between
open and closed string amplitudes, but they remain completely
obscure at the level of the Einstein-Hilbert
Lagrangian~\cite{Bern:1999ji,Ananth:2007zy}.

In this paper we present an algorithm for
writing down an arbitrary tree-level
SUGRA amplitude.  Our result was largely made possible by combining
and extending the results of two recent papers.
In~\cite{Drummond:2008cr}  an explicit
formula for all tree amplitudes in SYM was found by solving the supersymmetric
version~\cite{Brandhuber:2008pf,ArkaniHamed:2008gz} of the on-shell
recursion relation~\cite{Britto:2004ap,Britto:2005fq}, greatly extending
an earlier solution~\cite{Britto:2005dg} for split-helicity amplitudes
only.
We will review all appropriate details in a moment, but for now
it suffices to write their formula for the
color-ordered SYM amplitude $A(1,\ldots,n)$ very schematically as
\begin{equation}
\label{eqone}
A(1,\ldots,n) = A^{\rm MHV}(1,\ldots,n) \sum_{\{\alpha\}}
R_\alpha(\lambda_i, \widetilde{\lambda}_i, \eta_i)\,,
\end{equation}
where the sum runs over a collection of dual
superconformal~\cite{Drummond:2006rz,Drummond:2007aua,Drummond:2008vq}
invariants
$R_\alpha$.  The set $\{ \alpha \}$ is dictated by whether $A$
is MHV (in which case there is obviously only a single term, 1, in the sum),
next-to-MHV (NMHV), next-to-next-to-MHV (NNMHV), etc.

Our second inspiration is an intriguing formula for the $n$-graviton
MHV amplitude obtained by Elvang and Freedman~\cite{Elvang:2007sg}
which has the feature of expressing the amplitude in terms of sums of
squares of gluon amplitudes, in spirit similar to though in detail
very different
from the KLT relations.  Their formula reads
\begin{equation}
\label{eqtwo}
{\cal M}^{\rm MHV}_n = \sum_{{\mathcal{P}}(2,\ldots,n-1)}
[A^{\rm MHV}(1,\ldots,n)]^2 \, G^{\rm MHV}(1,\ldots,n)\,,
\end{equation}
where the sum runs over all permutations of the labels $2$ through $n-1$
and $G^{\rm MHV}(1,\ldots,n)$ is
a particular `gravity factor' reviewed below.

Our result involves a natural
merger of~(\ref{eqone}) and~(\ref{eqtwo}), expressing
an arbitrary $n$-graviton super-amplitude
in the form
\begin{equation}
\label{ourresult}
{\mathcal{M}}_n = \sum_{{\mathcal{P}}(2,\ldots,n-1)}
[A^{\rm MHV}(1,\ldots,n)]^2 \sum_{ \{\alpha \} }
[R_\alpha(\lambda_i, \widetilde{\lambda}_i, \eta_i) ]^2
\, G_\alpha(\lambda_i, \widetilde{\lambda}_i)\,.
\end{equation}
Two important features worth pointing out are that the sum runs
over precisely the same set $\{ \alpha \}$ that appears in the SYM
case~(\ref{eqone}), rather than some kind of double sum as one might
have guessed, and that the `gravity dressing factors' $G_\alpha$ do not
depend on the fermionic coordinates $\eta_i^A$ of
the on-shell ${\mathcal{N}} = 8$
superspace.  All of the `super' structure of the amplitudes is
completely encoded in the same $R$-factors that appear
already in the SYM amplitudes.

We begin in the next section by reviewing some of the necessary tools
for carrying out our calculation.  In section III we
provide detailed
derivations
of explicit formulas for MHV, NMHV, and NNMHV amplitudes.
Finally in section IV we discuss the structure of the gravity
dressing factors $G_\alpha$ for more general graviton amplitudes.

\section{Setting up the Calculation}

\subsection{Supersymmetric Recursion}

We will use the supersymmetric
version~\cite{Brandhuber:2008pf,ArkaniHamed:2008gz}
of the on-shell recursion
relation~\cite{Britto:2004ap,Britto:2005fq}
\begin{equation}
\mathcal{M}_n
= \sum_{P} \int
\frac{d^8\eta}{P^2} \mathcal{M}_{\text{L}}(z_{P})
\mathcal{M}_{\text{R}}(z_{P})
\label{SBCF}
\end{equation}
where we follow the conventions of~\cite{Drummond:2008cr}
in choosing
the supersymmetry preserving shift
\begin{align}
\label{shift}
{\lambda}_{\widehat{1}}(z) &= \lambda_1 - z \lambda_n\,,\cr
\widetilde{\lambda}_{\overline{n}}(z) &= \widetilde{\lambda}_n +
z \widetilde{\lambda}_1\,,\cr
\eta_{\overline{n}}(z) &= \eta_n + z \eta_1\,,
\end{align}
so that the sum in~(\ref{SBCF})
runs over all factorization channels of ${\mathcal{M}}_n$ which
separate
particle $1$ and particle $n$ (into ${\mathcal{M}}_{\text{L}}$
and ${\mathcal{M}}_{\text{R}}$, respectively).
The value of the shift parameter
\begin{equation}
z_P
= \frac{P^2}{[ 1 | P | n\rangle}
\label{zP}
\end{equation}
is chosen so that the shifted intermediate momentum
\begin{equation}
\widehat{P}(z) = P + z \lambda_n \widetilde{\lambda}_1\,,
\qquad
P = - p_1 - \cdots =
\cdots + p_n
\end{equation}
goes on-shell at $z=z_P$.
The recursion relation~(\ref{SBCF}) can be seeded with the
fundamental 3-particle amplitudes~\cite{ArkaniHamed:2008gz}
\begin{equation}
\label{threepoint}
{\mathcal{M}}^{\overline{\rm MHV}}_3 =
\frac{ \delta^{(8)}(
\eta_1 \bra{2}{3} + \eta_2 \bra{3}{1} + \eta_3 \bra{1}{2}) }
{ (\bra{1}{2} \bra{2}{3} \bra{3}{1})^2 }\,,
\qquad
{\mathcal{M}}^{\rm MHV}_3 =
\frac{\delta^{(16)}(q)}{(\ket{1}{2} \ket{2}{3} \ket{3}{1})^2}\,.
\end{equation}

\subsection{Gravity Subamplitudes}

\begin{figure}
\includegraphics{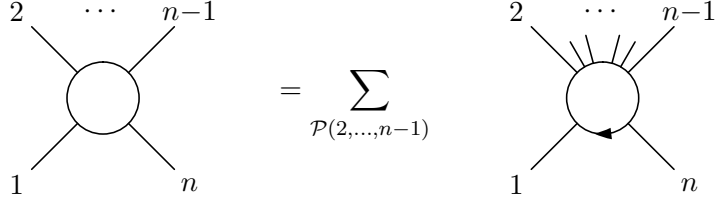}
\vskip -0.5cm
\caption{A diagrammatic representation of the relation~(\ref{main})
between
a physical gravity amplitude ${\mathcal{M}}_n$ and the sum over its
ordered subamplitudes $M(1,\ldots,n)$.
We draw an arrow indicating the cyclic order of the indices between
the special legs $n$ and $1$.}
\label{subamplitude}
\end{figure}

Color-ordered amplitudes in SYM have a cyclic structure such that
only those factorizations preserving the cyclic
labeling of the external particles appear in the analogous
recursion~(\ref{SBCF}).
In contrast, gravity amplitudes must be completely symmetric
under the exchange of any particle labels, so vastly more
factorizations contribute to~(\ref{SBCF}).
We can deal with this complication once and for all by introducing
the notion of an ordered `gravity subamplitude' $M(1,\ldots,n)$.
These
non-physical but mathematically useful
objects are related to the complete, physical
amplitudes ${\mathcal{M}}_n$ via the relation
\begin{equation}
\label{main}
{\mathcal{M}}_n = \sum_{{\mathcal{P}}(2,\ldots,n-1)}
M(1,\ldots,n)\,,
\end{equation}
depicted graphically in~\fig{subamplitude}.
This decomposition only makes a subgroup of the full permutation
symmetry manifest.
However it is the largest subgroup that the recursion~(\ref{SBCF}) allows
us to preserve since two external lines are singled out for special treatment.

The relation~(\ref{main}) does not uniquely determine the subamplitudes
for a given ${\mathcal{M}}_n$, since one could add to $M(1,\ldots,n)$
any quantity which vanishes after summing over permutations.
We choose to
define the subamplitudes $M$ recursively via~(\ref{SBCF})
restricted to factorizations which
preserve the cyclic ordering of the indices, just like in SYM theory:
\begin{equation}
\label{mdef}
M(1,\ldots,n) \equiv \sum_{i=3}^{n-1}
\int \frac{d^8\eta}{P^2}
M(\widehat{1},2,\ldots,i-1,\widehat{P})
M(-\widehat{P},i,\ldots,n-1,\overline{n})\,.
\end{equation}
This recursion is also seeded with the three-point
amplitudes~(\ref{threepoint})
since there is no distinction between $M(1,2,3)$ and ${\mathcal{M}}_3$.
Note however that unlike the color-ordered
SYM amplitudes $A(1,\ldots,n)$, the gravity subamplitude
$M(1,\ldots,n)$ is {\it not} in general
invariant under cyclic permutations of its
arguments.

It remains to prove the consistency of this definition.
That is, we need to check that the subamplitudes
defined in~(\ref{mdef}),
when substituted into~(\ref{main}),
do in fact give correct expressions for the physical gravity
amplitude ${\mathcal{M}}_n$.
This straightforward combinatorics exercise
proceeds by induction, beginning with the $n=3$ case which is trivial
and then
assuming that~(\ref{main}) is correct up to and including $n-1$
gravitons.
For $n$ gravitons we then have
\begin{align}
{\mathcal{M}}_n &= \sum_{A \bigcup B = \{2,\ldots,n-1\}}
\int \frac{d^8\eta}{P^2} {\mathcal{M}}(\widehat{1},\{A\},\widehat{P})
{\mathcal{M}}(-\widehat{P},\{B\},\overline{n})\cr
&= \frac{1}{(n-2)!} \sum_{{\mathcal{P}}(2,\ldots,n-1)}
\sum_{A \bigcup B = \{2,\ldots,n-1\}}
\int \frac{d^8\eta}{P^2} {\mathcal{M}}(\widehat{1},\{A\},\widehat{P})
{\mathcal{M}}(-\widehat{P},\{B\},\overline{n})
\cr
&= \frac{1}{(n-2)!} \sum_{{\mathcal{P}}(2,\ldots,n-1)}
\sum_{j=3}^{n-1} \binom{n-2}{j-2}
\int \frac{d^8\eta}{P^2} {\mathcal{M}}(\widehat{1},2,\ldots,j-1,\widehat{P})
{\mathcal{M}}(-\widehat{P},j,\ldots,n-1,\overline{n})
\cr
&=  \sum_{{\mathcal{P}}(2,\ldots,n-1)}
\sum_{j=3}^{n-1} \int \frac{d^8\eta}{P^2}
M(\widehat{1},2,\ldots,j-1,\widehat{P})
M(-\widehat{P},j,\ldots,n-1,\overline{n})
\cr
&= \sum_{{\mathcal{P}}(2,\ldots,n-1)}
M(1,2,\ldots,n)\,.
\end{align}
The first line is the superrecursion for the physical amplitude,
including a sum over all partitions of $\{2,\ldots,n-1\}$ into two
subsets $A$ and $B$,
not just those which preserve a cyclic
ordering.  In the second line we have thrown in a spurious
sum over all permutations of $\{2,\ldots,n-1\}$
at the cost of dividing
by $(n-2)!$ to
compensate for the overcounting.
This is allowed since we know that ${\mathcal{M}}_n$ is completely
symmetric under the exchange of any of its arguments.
Inside the sum over permutations we are then free to
choose $A = \{2,\ldots,i-1\}$ and $B = \{i,\ldots,n-1\}$ as indicated
on the third line, including the factor $\binom{n-2}{i-2}$ to
count the number of times this particular term appears.
On the fourth line our prior assumption that~(\ref{main})
holds up to $n-1$ particles allows us to replace
${\mathcal{M}}_a \to (a-2)! M_a$ inside the sum over permutations.
The last line invokes the
definition~(\ref{mdef}) and
completes the proof
that the physical $n$-graviton amplitude may be
recovered from the ordered subamplitudes via~(\ref{main}) and
the definition~(\ref{mdef}).

\subsection{From ${\mathcal{N}}=4$ to ${\mathcal{N}}=8$ Superspace}

The astute reader may have objected already to~(\ref{eqtwo})
in the introduction.
The SYM MHV amplitude~(\ref{AMHV})
involves the delta function $\delta^{(8)}(q)$ expressing conservation
of the total supermomentum
\begin{equation}
\label{qdef}
q =   \sum_{i=1}^n \lambda_i^\alpha
\eta_i^A\,, \qquad \alpha=1,2\,, \quad A=1,\ldots,4\,.
\end{equation}
Since the square of a fermionic delta function is zero,
it would seem that it makes no sense for the quantity
$[A^{\rm MHV}(1,\ldots,n)]^2$
to appear in~(\ref{eqtwo}).

Throughout this paper it will prove extremely convenient to adopt
the convention that the square of an ${\mathcal{N}}=4$ superspace
expression refers to an ${\mathcal{N}}=8$ superspace expression in
the most natural way.
For example, it should always be understood that
\begin{equation}
\label{convention}
[ \delta^{(8)} (q) ]^2 = \delta^{(16)}(q)\,,
\end{equation}
where the $q$ on the right-hand side is given by
the same expression~(\ref{qdef}) but with $A=1,\ldots,8$.
This notation will prove especially useful for lifting
results of Grassmann integration from ${\mathcal{N}} = 4$ to
${\mathcal{N}} = 8$ superspace.
This trick works because we can break the ${\rm SU}(8)$ symmetry
of a $d^8\eta$ integration
into ${\rm SU}(4)_a \times {\rm SU}(4)_b$ by taking
$\eta_1,\ldots,\eta_4$ for ${\rm SU}(4)_a$ and $\eta_5,\ldots,\eta_8$
for ${\rm SU}(4)_b$.  Then every $d^8 \eta$ integral can be rewritten
as a product of two SYM integrals and the ${\rm SU}(8)$ symmetry
of the answer is restored simply by adopting the
convention~(\ref{convention}).

For a specific example consider
the basic SYM integral
\begin{equation}
\int \frac{d^4 \eta}{P^2} \,A^{\overline{\rm MHV}}(\widehat{1},2,\widehat{P})
A^{\rm MHV}(-\widehat{P},3,\ldots,\overline{n})
= \frac{\delta^{(8)}(q)}{\ket{1}{2} \ket{2}{3} \cdots \ket{n}{1}}
\end{equation}
which expresses the superrecursion for the case of MHV amplitudes.
By `squaring' this formula
we immediately obtain the answer for a similar ${\mathcal{N}} = 8$
Grassmann integral,
\begin{equation}
\label{integralexample}
\int \frac{d^8 \eta}{P^2} \,
[A^{\overline{\rm MHV}}(\widehat{1},2,\widehat{P})]^2
[A^{\rm MHV}(-\widehat{P},3,\ldots,\overline{n})]^2
= P^2 \frac{\delta^{(16)}(p)}
{(\ket{1}{2} \ket{2}{3} \cdots \ket{n}{1})^2}\,.
\end{equation}
Note the extra factor of $P^2$ which appears on the right-hand
side because we have, for obvious reasons, chosen not to square the
propagator $1/P^2$ on the left.

\subsection{Review of SYM Amplitudes}

\label{SYMReview}

Given the above considerations it should come as no surprise that
we will be able to import much of the structure of SYM amplitudes
directly into our SUGRA results.  Therefore we
now review the results of~\cite{Drummond:2008cr} for
tree amplitudes in SYM.
Here and in all that follows we use the standard
dual superconformal~\cite{Drummond:2006rz,Drummond:2007aua,Drummond:2008vq}
notation
\begin{align}
x_{i j} &= p_i + p_{i+1} + \cdots + p_{j - 1}\,, \cr
\theta_{ij} &= \lambda_i \eta_i + \cdots + \lambda_{j-1} \eta_{j-1}\,,
\end{align}
where all subscripts are understood mod $n$.

We will base our expression for the SUGRA amplitudes on an expression
for the SYM amplitudes
which is equivalent to, but not exactly the same as the one presented
in~\cite{Drummond:2008cr}.
The reason is that the cyclic symmetry of the Yang-Mills amplitudes
implies certain identities for the invariants $R_\alpha$ appearing
in~(\ref{eqone}). This symmetry was used
in~\cite{Drummond:2008cr} when solving the recursion relations.
Instead it is helpful to have a different
expression which is more suitable to the gravity case where the
subamplitudes $M$ do not have cyclic symmetry.

To be precise we need to return to the construction of~\cite{Drummond:2008cr}
and make sure that when considering the right-hand
side of the BCF recursion relation we
always insert the lower point amplitudes so that leg 1 of the left amplitude
factor corresponds to the shifted leg $\widehat{1}$. We also need to have the
leg $n$ of the right amplitude factor corresponding to the shifted leg
$\overline{n}$, but this was already the choice made in~\cite{Drummond:2008cr}.

\begin{figure}
\psfrag{one}[cc][cc]{$1$}
\psfrag{a1b1}[cc][cc]{$a_{1}b_{1}$}
\psfrag{a2b2}[cc][cc]{$a_{2}b_{2}$}
\psfrag{a3b3}[cc][cc]{$a_{3}b_{3}$}
\psfrag{b1a1a2b2}[cc][cc]{$a_{1}b_{1};a_{2}b_{2}$}
\psfrag{b2a2a3b3}[cc][cc]{$a_{2}b_{2};a_{3}b_{3}$}
\psfrag{b1a1a3b3}[cc][cc]{$a_{1}b_{1};a_{3}b_{3}$}
\psfrag{b1a1b2a2a3b3}[cc][cc]{$a_{1}b_{1};a_{2}b_{2};a_{3}b_{3}$}
\psfrag{two}[cc][cc]{$2$}
\psfrag{n1}[cc][cc]{$n$}
\psfrag{a1p}[cc][cc]{$a_{1}$}
\psfrag{a2p}[cc][cc]{$a_{2}$}
\psfrag{b1}[cc][cc]{$b_{1}$}
\psfrag{b2}[cc][cc]{$b_{2}$}
\centerline{{\epsfysize7cm
\epsfbox{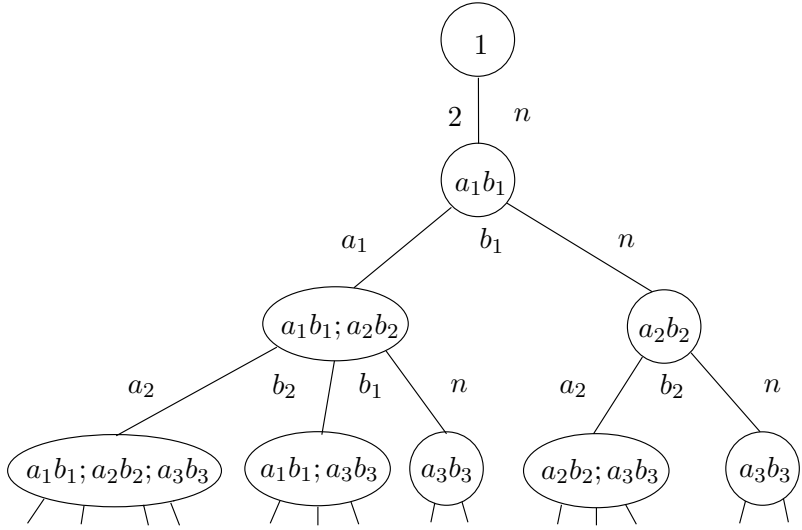}}}
\caption{An alternative rooted tree diagram for
tree-level SYM amplitudes. The figure is the same as the tree diagram
presented in \cite{Drummond:2008cr} except that the labels in the vertices
appear in a different order, meaning that the $R$-invariants appearing
in the amplitude are slightly different. Also the limits, written to the
left and right of each line, are treated differently.}
\label{newYMtree}
\end{figure}

The expression for all $\mathcal{N}=4$ SYM amplitudes is given in terms of
paths in a particular rooted tree diagram. Here we will be using a different
(but equivalent) diagram, shown in~\fig{newYMtree}. Each vertex in the diagram,
say with labels $a_1b_1;a_2b_2;\ldots;a_rb_r;ab$, corresponds to a particular
dual conformal invariant. These invariants take the general form \cite{Drummond:2008vq,Drummond:2008cr}
\begin{equation}
R_{n;a_1b_1;a_2b_2;\ldots;a_rb_r;ab} =
\frac{\ket{a}{a-1} \ket{b}{b-1}
\ \delta^{(4)}(\langle \xi | x_{b_r a}x_{ab} | \theta_{b b_r} \rangle +
\langle \xi |x_{b_r b} x_{ba} | \theta_{ab_r} \rangle)}
{x_{ab}^2 \langle \xi |
x_{b_r a} x_{ab} | b \rangle \langle \xi | x_{b_r a} x_{ab} |b-1\rangle
\langle \xi
| x_{b_r b} x_{ba} |a\rangle \langle \xi | x_{b_r b} x_{ba} |a-1\rangle}\,,
\label{generalR}
\end{equation}
where the chiral spinor $\xi$ is given by
\begin{equation}
\langle \xi | =
\langle n | x_{na_1}x_{a_1 b_1} x_{b_1 a_2} x_{a_2 b_2} \ldots x_{a_r b_r} \,.
\end{equation}
As in~\cite{Drummond:2008cr} this expression
needs to be slightly modified when any $a_i$ index
attains the lower limit of its range\footnote{
In~\cite{Drummond:2008cr} it was also necessary to sometimes take into account
modifications when indices reached the upper limits of their ranges,
but this feature does not arise in our reorganized
presentation of the amplitude.}.
We indicate
by means of a superscript on $R$ the nature of the
appropriate modification.
Specifically,
$R_{n;a_1b_1;a_2b_2;\ldots;a_rb_r;ab}^{l_1,\ldots,l_r}$
indicates the same quantity~(\ref{generalR}) but with the understanding
that when $a$ reaches its lower limit, we need to replace
\begin{equation}
\langle a{-}1 | \to \langle n| x_{nl_1} x_{l_1l_2} \cdots x_{l_{r-1} l_r}\,.
\end{equation}

We now have all of the ingredients necessary to begin assembling
the complete amplitude, which is given by the formula
\begin{equation}
A_n = A_n^{\rm MHV} \mathcal{P}_n =
\frac{ \delta^{(8)}(q)}{\ket{1}{2} \cdots \ket{n}{1}}
\mathcal{P}_n\,,
\end{equation}
where ${\mathcal{P}}_n$ is given by the sum over vertical paths
in~\fig{newYMtree} beginning at the root node.
To each such path
we associate a nested sum of the product of the associated $R$-invariants
in the vertices visited by the path.
The last pair of labels in a given $R$ are those which are summed
first, these are denoted by $a_p b_p$ in row $p$ of the diagram.
We always take the convention that $a_p$ and $b_p$ are separated
by at least two ($a_p < b_p - 1$) which is necessary for the $R$-invariants
to be well-defined.
The lower and upper limits
for the summation variables $a_p,b_p$ are indicated by the two numbers
appearing adjacent to the line above each vertex.

\begin{figure}
\psfrag{uvs}[cc][cc]{$u_{1}v_{1};\ldots u_{r}v_{r};a_{p-1}b_{p-1}$}
\psfrag{uvab}[cc][cc]{$u_{1}v_{1};\ldots u_{r}v_{r};a_{p-1}b_{p-1};a_{p}b_{p}$}
\psfrag{uvnext}[cc][cc]{$u_{1}v_{1};\ldots u_{r}v_{r};a_{p}b_{p}$}
\psfrag{ab}[cc][cc]{$a_{p}b_{p}$}
\psfrag{up}[cc][cc]{$a_{p-1}$}
\psfrag{bp1}[cc][cc]{$b_{p-1}$}
\psfrag{bvr}[cc][cc]{$v_{r}$}
\psfrag{bv1}[cc][cc]{$v_{1}$}
\psfrag{n1}[cc][cc]{$n$}
\psfrag{dots}[cc][cc]{$\ldots$}
\centerline{{\epsfysize4cm
\epsfbox{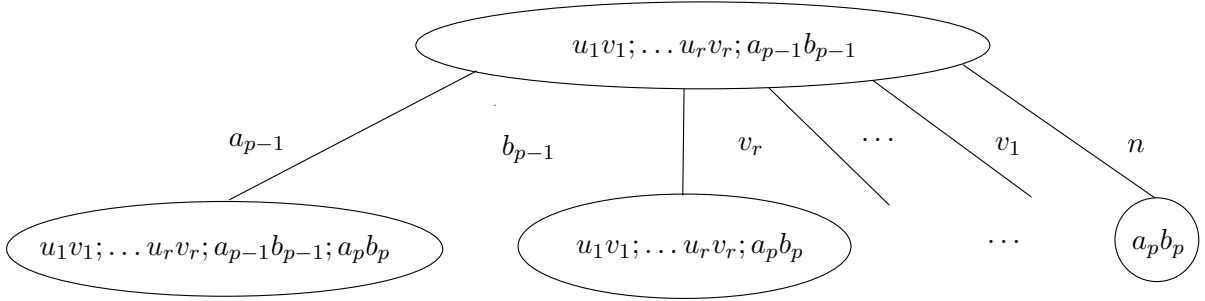}}}  \caption[]{
The rule for going from line $p-1$ to line $p$ (for $p>1$)
in~\fig{newYMtree}. For every vertex in line $p-1$ of the form given at the
top of the diagram, there are $r+2$ vertices in the lower line (line $p$).
The labels in these vertices start with
$u_{1}v_{1};\ldots u_{r}v_{r};a_{p-1}b_{p-1};a_{p}b_{p}$ and they get
sequentially shorter, with each step to the right removing the pair of
labels adjacent to the last pair $a_p,b_p$ until only the last pair is left.
The summation limits between each line are also derived from the labels
of the vertex above.
The left superscripts which appear on the associated $R$-invariants start
with $u_1v_1\ldots u_rv_r b_{p-1} a_{p-1}$ for the left-most vertex. The
next vertex to the right has the superscript
$u_1v_1 \ldots u_r v_r a_{p-1} b_{p-1}$, i.e. the same as the first but
with the final pair in alphabetical order. The next vertex has the
superscript $u_1v_1 \dots u_rv_r$ and thereafter the pairs are sequentially
deleted from the right.}
\label{cluster}
\end{figure}

The differences between
the new diagram and the one of~\cite{Drummond:2008cr} are:

\begin{enumerate}
\item{ All pairs of labels in the vertices
appear alphabetically in the form $a_i b_i$.}

\item{The
edges on the extreme left of the diagram are labeled by $a_i$ rather
than $a_i + 1$, and
the summation variables must be greater than or equal to these lower
limits $a_i$.}

\item{The edges on the extreme right of the
diagram are labelled by $n$ rather than $n-1$,  and the summation
variables must be strictly less than this upper limit $n$.}

\item{ All superscripts on $R$-invariants which detail boundary
replacements are left superscripts (i.e. for lower boundaries only).
In a given cluster, e.g. the cluster shown
in~\fig{cluster},  the superscript associated to the left-most vertex
is obtained from the sequence written in the vertex by deleting the final
pair of labels and reversing the order of the last two labels which remain.
Thus the sequence ends $b_i a_i$ for some $i$. Then proceeding to the
right in the cluster, the next vertex has the same superscript, but with
alphabetical order of the final pair, i.e. it ends $a_i b_i$. Going
further to the right in the cluster one obtains the relevant
superscripts by sequentially deleting pairs of labels from the right.}
\end{enumerate}

Given the complexity of this prescription it behooves us to
illustrate a few cases explicitly.
There is one path of
length zero, whose value is simply 1 and this corresponds to the MHV
amplitudes,
\begin{equation}
\mathcal{P}_n^{\rm MHV} = 1\,.
\end{equation}
Then there is one path of length one which gives the NMHV amplitudes.
We get $1 \times R_{n;a_1,b_1}$, summed over the region $2\leq a_1, b_1 < n$,
as always with the convention that $a_i < b_i-1$. There are no boundary
replacements so we have
\begin{equation}
\mathcal{P}_n^{\rm NMHV} =
\sum_{2\leq a_1,b_1 < n} \!\!\!\!\!\! R_{n;a_1b_1} \, .
\end{equation}
The two paths of length two give the NNMHV amplitudes. This time we get
superscripts on the $R$-invariants as dictated by the rules in point 4
above,
\begin{equation}
\mathcal{P}_n^{\rm  NNMHV} =
\sum_{2 \leq a_1, b_1 < n} \!\!\!\!\!\! R_{n;a_1b_1}
\Big( \sum_{a_1 \leq a_2,b_2 < b_1} \!\!\!\!\!\!
R_{n;a_1b_1;a_2b_2}^{b_1a_1} +
\sum_{b_1 \leq a_2,b_2 < n} \!\!\!\!\!\! R_{n;a_2b_2}^{a_1b_1} \Big) \, .
\end{equation}
Continuing to N${}^3$MHV amplitudes we find five paths of length three,
giving the following nested sums,
\begin{align}
&\mathcal{P}_n^{{\rm N}^3{\rm MHV}} =
\sum_{2 \leq a_1, b_1 < n} \!\!\!\!\!\! R_{n;a_1b_1} \Bigl[  \nonumber \\
&\sum_{a_1 \leq a_2,b_2 < b_1} \!\!\!\!\!\! R_{n;a_1b_1;a_2b_2}^{b_1a_1}
\Bigl(\sum_{a_2 \leq a_3,b_3 < b_2} \!\!\!\!\!\!
R_{n;a_1b_1;a_2b_2;a_3b_3}^{a_1b_1b_2a_2} +
\sum_{b_2 \leq a_3,b_3 < b_1} \!\!\!\!\!\!
R_{n;a_1b_1;a_3b_3}^{a_1b_1a_2b_2} +
\sum_{b_1 \leq a_3,b_3 < n} \!\!\!\!\!\!
R_{n;a_3b_3}^{a_1b_1} \Bigr) \notag \\
&+ \sum_{b_1 \leq a_2,b_2 < n} \!\!\!\!\!\!
R_{n;a_2b_2}^{a_1b_1} \Bigl(\sum_{a_2 \leq a_3,b_3 < b_2} \!\!\!\!\!\!
R_{n;a_2b_2;a_3b_3}^{b_2a_2}
+ \sum_{b_2 \leq a_3,b_3 < n} \!\!\!\!\!\! R_{n;a_3b_3}^{a_2b_2} \Bigr)
\Bigr] \, .
\end{align}
These examples hopefully serve to illustrate how to write
a general SYM amplitude, though a more thorough
discussion may be found in~\cite{Drummond:2008cr}.

\section{Examples of Gravity Amplitudes}

\subsection{MHV Amplitudes}

Elvang and Freedman have
shown
that the $n$-graviton MHV amplitude may be
written in the
form\footnote{
We have relabeled their indices according to $i \to 2 - i~{\rm mod}~n$ and
have expressed the amplitude in ${\mathcal{N}} = 8$ superspace.}
\begin{equation}
\label{EFMHV}
{\mathcal{M}}^{\rm MHV}_n = \sum_{{\mathcal{P}}(2,\ldots,n-1)}
[A^{\rm MHV}(1,\ldots,n)]^2
G^{\rm MHV}(1,\ldots,n)
\end{equation}
in terms of
\begin{equation}
\label{MHVG}
G^{\rm MHV}(1,\ldots,n) =
x_{13}^2 \prod_{s=2}^{n-3}
\frac{\langle s | x_{s,s+2} x_{s+2,n} |n\rangle}{\ket{s}{n}}\,.
\end{equation}
The formula~(\ref{EFMHV}) is valid for $n > 3$; $n=3$ will always
be treated as a special case with $G^{\rm MHV}(1,2,3) = 1$.

Comparison of~(\ref{EFMHV}) with~(\ref{main}) suggests that
we should identify the MHV ordered subamplitude as
\begin{equation}
\label{subMHV}
M^{\rm MHV}(1,\ldots,n) = [A^{\rm MHV}(1,\ldots,n)]^2
G^{\rm MHV}(1,\ldots,n)\,.
\end{equation}
Let us now check that our definition~(\ref{mdef}) yields
precisely the same expression for the subamplitude
(they may have differed by terms which cancel out when one sums over
all permutations in~(\ref{main})).

\begin{figure}
\includegraphics{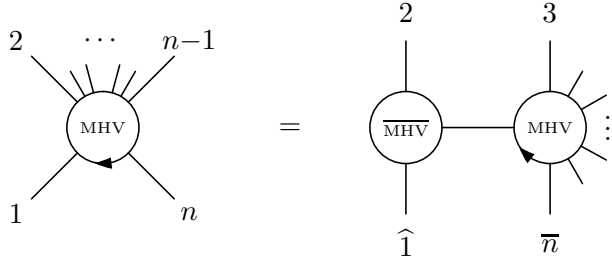}
\vskip -0.5cm
\caption{The recursion for MHV amplitudes.}
\label{fig:mhv}
\end{figure}

We will again proceed by induction, assuming that~(\ref{subMHV})
satisfies~(\ref{mdef}) for $n-1$ and fewer gravitons.
To calculate $M^{\rm MHV}$ for $n$ gravitons
from the definition~(\ref{mdef})
we first note that only the single term $i=3$ contributes,
giving
\begin{equation}
\label{MHVBCF}
M^{\rm MHV}(1,\ldots,n) =
\int \frac{d^8 \eta}{P^2} \,M^{\overline{\rm MHV}}(\widehat{1},2,\widehat{P})
M^{\rm MHV}(-\widehat{P},3,\ldots,\overline{n})
\end{equation}
as shown in~\fig{fig:mhv}.
The calculation is rendered essentially trivial by plugging in the relations
\begin{align}
M^{\overline{\rm MHV}}
(\widehat{1},2,\widehat{P}) &=
[A^{\overline{\rm MHV}}(\widehat{1},2,\widehat{P})]^2\,,
\cr
M^{\rm MHV}(-\widehat{P},3,\ldots,\overline{n})
&= [A^{\rm MHV}(-\widehat{P},3,\ldots,\overline{n})]^2
\,G^{\rm MHV}(-\widehat{P},3,\ldots,\overline{n})
\end{align}
between ordered graviton and Yang-Mills amplitudes.
The $G$ factor in~(\ref{MHVBCF}) comes along for the ride
as we perform the $d^8 \eta$ integral using the square
of the analogous Yang-Mills calculation as explained
above~(\ref{integralexample}).
Therefore with no effort we find that~(\ref{MHVBCF}) gives
\begin{equation}
M^{\rm MHV}(1,\ldots,n) =  [A^{\rm MHV}(1,\ldots,n)]^2
\, P^2
G^{\rm MHV}(-\widehat{P},3,\ldots,\overline{n})\,.
\end{equation}
A simple calculation using the shift~(\ref{shift})
now reveals that
\begin{align}
P^2 G^{\rm MHV}(-\widehat{P},3,\ldots,\overline{n})
&= x_{13}^2 (-\widehat{P} + p_3)^2 \prod_{s=3}^{n-3}
\frac{\langle s| x_{s,s+2} x_{s+2,\overline{n}}|\overline{n}\rangle}{
\ket{s}{n}}
\cr
&= x_{13}^2 \prod_{s=2}^{n-3}
\frac{\langle s |x_{s,s+2} x_{s+2,n} |n\rangle
}{\ket{s}{n}}
\cr
&= G^{\rm MHV}(1,\ldots,n)\,.
\end{align}
This completes the inductive
proof that the formula~(\ref{subMHV}) obtained by Elvang
and Freedman
is precisely the MHV case of the ordered subamplitudes that we have
defined
in~(\ref{mdef}).

\subsection{NMHV Amplitudes}

\begin{figure}
\includegraphics{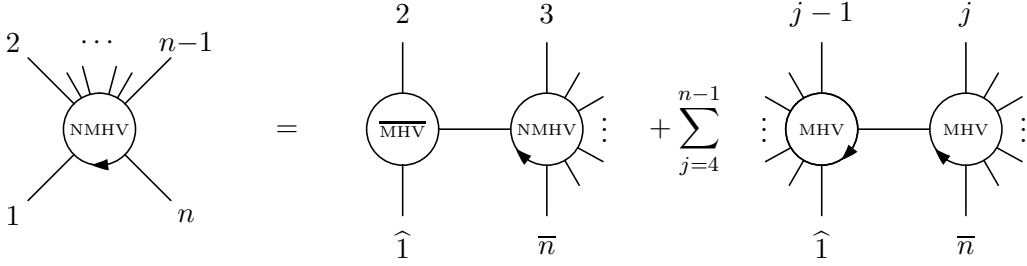}
\vskip -0.5cm
\caption{The two kinds
of diagrams contributing to the recursion of NMHV amplitudes.
}
\label{fig:nmhv}
\end{figure}

Next we turn our attention to the NMHV amplitude.
The two kinds of diagrams which contribute to the recursion
are shown in~\fig{fig:nmhv}.
Let us begin with $n=5$, in which case the first
diagram is absent and only the term $i=4$ appears in the sum.
According to the
definition~(\ref{mdef}) we then have
\begin{align}
M^{\rm NMHV}(1,\ldots,5) &= \int \frac{d^8 \eta}{P^2}
M^{\rm MHV}(\widehat{1},2,3,\widehat{P})
M^{\rm MHV}(-\widehat{P},4,\overline{5}) \cr
&= [A^{\rm NMHV}(1,\ldots,5)]^2 \,P^2
\,
G^{\rm MHV}(\widehat{1},2,3,\widehat{P})\cr
&\equiv
[A^{\rm NMHV}(1,\ldots,5)]^2 \,
G^{\rm NMHV}(1,\ldots,5)\,.
\end{align}
Here, following the example set in the previous
subsection, evaluating the Grassmann integral
leads to the square of the analogous SYM result, times
the gravity factor
\begin{equation}
G^{\rm NMHV}(1,\ldots,5)
=P^2 G^{\rm MHV}(\widehat{1},2,3,\widehat{P}) = (p_4 + p_5)^2
(p_{\widehat{1}} + p_2)^2 = (p_4 + p_5)^2
\frac{[4|p_3 p_2|1]}{\bra{4}{1}}\,.
\end{equation}
One can check that this result it is consistent with the known answer
(for example, from the KLT relation).

Let us now turn to the general NMHV case.
In the previous section we recalled the SYM result
obtained in~\cite{Drummond:2008cr},
\begin{equation}
\label{DHNMHV}
A^{\rm NMHV}(1,\ldots,n) = A^{\rm MHV}(1,\ldots,n) \sum_{i=2}^{n-3}
\sum_{j=i+2}^{n-1} R_{n;ij}\,.
\end{equation}
It was shown in~\cite{Drummond:2008cr} that the $i=2$
term in~(\ref{DHNMHV}) corresponds to the sum over
${\rm MHV} \times
{\rm MHV}$ diagrams in~\fig{fig:nmhv}, while the $i>2$ terms
arise iteratively from the $\overline{\rm MHV}
\times {\rm NMHV}$ diagram.

\subsubsection{Statement}

Now we claim that the NMHV gravity subamplitude is
given by
\begin{equation}
\label{NMHVclaim}
M^{\rm NMHV}(1,\ldots,n) = [A^{\rm MHV}(1,\ldots,n)]^2
\sum_{i = 2}^{n-3} \sum_{j=i+2}^{n - 1} R_{n;ij}^2 G^{\rm NMHV}_{n;ij}\,
\end{equation}
where $R$ is the same dual superconformal invariant~(\ref{generalR})
as in SYM and the NMHV gravity factor
can be split for future convenience into three parts as follows,
\begin{equation}
G^{\rm NMHV}_{n;ab} = f_{n;ab} G^L_{n;ab} G^R_{n;ab}\,.
\label{GNMHV}
\end{equation}
To express the gravity factor we introduce the notation
\begin{align}
P^{l,u}_{a_1,\ldots,a_r} &= \prod_{k=l}^u
\frac{\langle k | x_{k,k+2}x_{k+2,a_1} x_{a_1 a_2} x_{a_2 a_3} \ldots
x_{a_{r-1} a_r} |a_r \rangle}{\langle k | x_{a_1 a_2} x_{a_2 a_3}
\ldots x_{a_{r-1} a_r} | a_r \rangle}\,,\\
Z^{a_1,\ldots,a_u}_{b_1,\ldots,b_l ; c_1,\ldots,c_r} &=
\frac{\langle a_1 | x_{a_1 a_2} x_{a_2 a_3}\dots x_{a_{u-1} a_u} | a_u \rangle}
{\langle b_1 | x_{b_1 b_2} x_{b_2 b_3} \ldots
x_{b_{l-1} b_l} x_{c_1 c_2} x_{c_2 c_3} \ldots x_{c_{r-1} c_r} |c_r \rangle}\,,
\end{align}
which is overkill at the moment but will be fully utilized below
when we move beyond
the NMHV level.
In the numerators only dual conformal chains of $x$-matrices appear, while in
the denominators the chains are not dual conformal due to the break in the
way the labels are arranged. The break is denoted by the semi-colon in the
subscript of $Z$ while in the denominator of $P$ it is immediately after
the left-most spinor $\langle k |$.

Then the first factor in~(\ref{GNMHV}) is given by
\begin{align}
f_{n;2b} &= x_{1b}^2\, \label{f1}\,,\\
f_{n;ab} &= x_{13}^2 (-Z^{n,b,a-1}_{n;a-1}) P_n^{2,a-2}\qquad
\text{ for } a>2\,,
\label{f2}
\end{align}
while the remaining two are
\begin{align}
G^L_{n;ab} &= -Z^{n,a+1,b,a,n}_{n;b,a,n} P^{a,b-3}_{b,a,n}\, ,\label{GL}\\
G^R_{n;ab} &= -Z^{n,b+1,b,a,n}_{n;b,a,n} P^{b,n-3}_n \, . \label{GR}
\end{align}

\subsubsection{Proof}

To check that the formula~(\ref{NMHVclaim})
is correct it is useful to first have a general formula
for $x_{\widehat{1} v}^2$, where the shift is defined so that
$\widehat{P}_i^2=x_{\widehat{1} i}^2=0$. This tells us that the shift parameter
is given by~(\ref{zP}), i.e
\begin{equation}
z_P = \frac{x_{1i}^2}{\langle n | x_{1i} | 1]}\,.
\end{equation}
Then we have
\begin{align}
x_{\widehat{1} v}^2 &= x_{1v}^2 - z_P \langle n| x_{1v} | 1] \\
&= \frac{x_{1v}^2 \langle n | x_{1i} | 1] - x_{1i}^2 \langle n | x_{1v} | 1]}
{\langle n | x_{1i} | 1]} \\
&= \frac{\langle n | x_{1v} (x_{1v} - x_{1i})  x_{1i}|1]}
{\langle n | x_{1i} | 1]}
\label{firstchoice} \\
&= \frac{\langle n | x_{1v} x_{iv} x_{1i} | 1]}{\langle n | x_{1i} | 1]} \\
&= -\frac{ \langle n | x_{nv} x_{vi} x_{i2} x_{2n} | n \rangle}
{\langle n | x_{i2} x_{2n} | n \rangle} \equiv - Z^{n,v,i,2,n}_{n;i,2,n}\,.
\end{align}
Note that instead of writing~(\ref{firstchoice})
we could have alternatively written it as
\begin{align}
x_{\widehat{1} v}^2 &=\frac{\langle n | x_{1i} (x_{1v} - x_{1i})  x_{1v}|1]}
{\langle n | x_{1i} | 1]} \\
&= \frac{\langle n | x_{1i} x_{iv} x_{1v}|1]}{\langle n | x_{1i} | 1]} \\
&= \frac{\langle n | x_{ni} x_{iv} x_{v2} x_{2n} | n \rangle}
{\langle n | x_{i2} x_{2n} | n \rangle} \equiv Z^{n,i,v,2,n}_{n;i,2,n}\,.
\end{align}
The freedom to write this factor in these two various forms is useful
because in certain cases either one or the other form simplifies by
cancelling factors from the numerator and denominator.

Finally we are set up to check our claim~(\ref{GNMHV})
for the NMHV $G$-factor.  We first check the
case $a=2$ which comes entirely from MHV $\times$ MHV diagrams. From
these diagrams we obtain
\begin{equation}
\sum_{i=4}^{n-1} R^2_{n;2,i} G^{\rm NMHV}_{n;2,i} =
\sum_{i=4}^{n-1} R_{n;2,i}^2
P^2 G^{\rm MHV}(\widehat{1}, \ldots ,-\widehat{P})
G^{\rm MHV}(\widehat{P},\ldots,\overline{n})\,,
\end{equation}
from which we find
\begin{align}
G^{\rm NMHV}_{n;2,i} &= x_{1i}^2 \Bigl( x_{\widehat{1} 3}^2 \prod_{k=2}^{i-3}
\frac{\langle k | x_{k, k+2} x_{k+2,i} | \widehat{P} \rangle}
{\ket{k}{\widehat{P}}} \Bigr)\Bigl(x_{\widehat{1} i+1}^2
\prod_{l=i}^{n-3}
\frac{\langle l | x_{l,l+2}x_{l+2,n} | n \rangle}
{\ket{l}{n}}\Bigr)\\
&= x_{1i}^2 \bigl( -Z^{n,3,i,2,n}_{n;i,2,n} P^{2,i-3}_{i,2,n} \bigr)
\bigl( - Z^{n,i+1,i,2,n}_{n;i,2,n} P^{i,n-3}_n \bigr)\,,
\end{align}
which is in agreement with equations~(\ref{GNMHV}) to~(\ref{GR}) for the
case $a=2$.

For the case $a>2$ we must consider diagrams of the form
$\overline{\rm MHV}_3 \times {\rm NMHV}_{n-1}$. From these diagrams we obtain
\begin{equation}
\label{twoterms}
\sum_{3\leq a,b \leq n-1} R^{2}_{n;ab} G^{\rm NMHV}_{n;ab} =
\sum_{3\leq a,b \leq n-1}
R^{2}_{n;ab} P^2 G^{\rm NMHV}(\widehat{P},3,\ldots,\overline{n})\,.
\end{equation}
The sum splits into two contributions, $a=3$ and $a>3$. The first gives
\begin{align}
G^{\rm NMHV}_{n;3b} &= x_{13}^2 x_{\widehat{1} b}^2 \bigl(-Z^{n,4,b,3,n}_{n;b,3,n}
P^{a,b-3}_{b,a,n} \bigr) \bigl(-Z^{n,b+1,b,3,n}_{n;b,3,n} P^{b,n-3}_n \bigr)
\, \label{a=3first}\\
&= x_{13}^2 \bigl(- Z^{n,b,2}_{n;2} \bigr) \bigl(-Z^{n,4,b,3,n}_{n;b,3,n}
P^{a,b-3}_{b,a,n} \bigr) \bigl(-Z^{n,b+1,b,3,n}_{n;b,3,n} P^{b,n-3}_n \bigr)\,,
\label{a=3second}
\end{align}
in agreement with equations~(\ref{GNMHV}) to~(\ref{GR})
for the case $a=3$.
To go from~(\ref{a=3first}) to~(\ref{a=3second}) we have used the fact
that $x_{\widehat{1} b}^2 = -Z^{n,b,3,2,n}_{n;3,2,n} = - Z^{n,b,2}_{n;2}$
where the simplification of the $Z$-factor is due to a cancellation between
its numerator and denominator.

For the contributions to~(\ref{twoterms}) where $a>3$ we find
\begin{align}
G^{\rm NMHV}_{n;ab} &=
x_{13}^2 x_{\widehat{1} 4}^2 \bigl(-Z^{n,b,a-1}_{n;a-1}\bigr) P^{3,a-2}_n
\bigl(-Z^{n,a+1,b,a,n}_{n;b,a,n} P^{a,b-3}_{b,a,n} \bigr)
\bigl( -Z^{n,b+1,b,a,n}_{n;b,a,n} P^{b,n-3}_n \bigr) \, \\
&= x_{13}^2 \bigl(-Z^{n,b,a-1}_{n;a-1}\bigr) P^{2,a-2}_n
\bigl(-Z^{n,a+1,b,a,n}_{n;b,a,n} P^{a,b-3}_{b,a,n} \bigr)
\bigl( -Z^{n,b+1,b,a,n}_{n;b,a,n} P^{b,n-3}_n \bigr)\,,
\end{align}
which is again in agreement with equations~(\ref{GNMHV}) to~(\ref{GR}).
The factor $x_{\widehat{1}4}^2$ completes the factor $P^{3,a-2}_n$ to
$P^{2,a-2}_n$ just as in the MHV case.
This completes the verification of
the formula~(\ref{NMHVclaim}) for NMHV graviton amplitudes.
Appendix B contains some notes on extracting NMHV graviton amplitudes
from the super-amplitude~(\ref{NMHVclaim}).

\subsection{NNMHV Amplitudes}

In this section we consider the NNMHV case as an exercise towards
finding the general algorithm for all tree-level gravity amplitudes.

\subsubsection{Statement}

The structure of the result is just like in Yang-Mills and similar
to the NMHV case~(\ref{NMHVclaim}) except
that we now have two more subscripts on both the Yang-Mills $R$-factors
and the gravity factors,
\begin{equation}
\frac{M^{\rm NNMHV}(1,\ldots,n)}{
[A^{\rm MHV}(1,\ldots,n)]^2}=
\sum_{2\leq a,b \leq n-1} R_{n;a b}^2\Bigl[\sum_{a \leq c,d <
b} (R^{ba}_{n;ab;cd})^2 H^{(1)}_{n;ab;cd}
+ \sum_{b \leq c,d < n}
(R_{n;cd}^{ab})^2 H^{(2)}_{n;ab;cd} \Bigr]\, .
\label{MNNMHV}
\end{equation}
The factors $H^{(1)}$ and $H^{(2)}$ can be written in the form
\begin{align}
H^{(1)}_{n;ab;cd} &=
f_{n;ab} G^R_{n;ab} \widetilde{f}_{n;ab;cd} G^L_{n;ab;cd} G^R_{n;ab;cd} \,,
\label{H1}
\\
H^{(2)}_{n;ab;cd} &= f_{n;ab} G^L_{n;ab} \widehat{f}_{n;ab;cd}
G^L_{n;cd} G^R_{n;cd} \, .
\label{H2}
\end{align}
In this formula $f_{n;ab}$, $G^L_{n;ab}$ and $G^R_{n;ab}$ are defined as before in the
case of the NMHV amplitude (see formulae~(\ref{f2}),
(\ref{GL}) and~(\ref{GR})). The factor $\widetilde{f}$ in $H^{(1)}$ is given by
\begin{align}
\widetilde{f}_{n;ab,ad} &= -Z^{n,b,d,a,n}_{n;b,a,n}\,\label{factor f1} ,\\
\widetilde{f}_{n;ab;cd} &= \bigl(-Z^{n,b,a+1,a,n}_{n;b,a,n}
\bigr)\bigl(-Z^{c-1,d,b,a,n}_{c-1;b,a,n}\bigr) P^{a,c-2}_{b,a,n} \,
\qquad \text{ for } c>a\,,\label{factor f2}
\end{align}
and the factor $\widehat{f}$ in the second term in the parentheses is given by
\begin{align}
\widehat{f}_{n;ab;bd} &= - Z^{n,d,b,a,n}_{n;b,a,n} \, \\
\widehat{f}_{n;ab;cd} &= \bigl(-Z^{n,b+1,b,a,n}_{n;b,a,n} \bigr)
\bigl(-Z^{n,d,c-1}_{n;c-1} \bigr) P_n^{b,c-2} \qquad \text{ for } c > b\,.
\end{align}
Finally the new $G$-factors are given by
\begin{align}
{G}^L_{n;ab;cd}  &=
-Z^{n,a,b,c+1,d,c,b,a,n}_{n,a,b;d,c,b,a,n} P^{c,d-3}_{d,c,b,a,n} \, ,\\
{G}^R_{n;ab;cd} &=
-Z^{n,a,b,d+1,d,c,b,a,n}_{n,a,b;d,c,b,a,n} P^{d,n-3}_{b,a,n} \, .
\end{align}

\subsubsection{Proof}

Let us now
check the claim~(\ref{MNNMHV}). As before we begin with the case $a=2$ which
comes purely from NMHV $\times$ MHV diagrams and MHV $\times$ NMHV diagrams.
We start by calculating the former kind. From these diagrams we obtain
\begin{align}
&\sum_{i=5}^{n-1} R^2_{n;2i}
\sum_{2\leq c,d < i} \!\!\!
(R^{i2}_{n;2i;cd})^2 H^{(1)}_{n;2i,cd} \notag\\
=& \sum_{i=5}^{n-1} R^2_{n;2i}
\sum_{2\leq c,d < i} \!\!\!
(R^{i2}_{n;2i;cd})^2 P^2 G^{\rm NMHV}(\widehat{1},\ldots,-\widehat{P})
G^{\rm MHV}(\widehat{P},\ldots,\overline{n})\,.
\end{align}

The sum over $c$ splits into two pieces, $c=2$ and $c>2$. For the terms
where $c=2$ we have
\begin{align}
H^{(1)}_{n;2i;2d} =
x_{1i}^2 \Bigl[x_{\widehat{1} d}^2 \bigl(-Z^{n,2,i,3,d,2,n}_{n,2,i;d,2,n}
P^{2,d-3}_{d,2,n} \bigr) \bigl(-Z^{n,2,i,d+1,d,2,n}_{n,2,i;d,2,n}
P^{d,i-3}_{i,2,n} \bigr)\Bigr] \Bigl[ x_{\widehat{1},i+1}^2 P^{i,n-3}_n\Bigr]\,.
\label{HNNMHVa=c=2}
\end{align}
Here as in the previous
subsection
we have used the fact that certain $Z$-factors simplify. For example,
reading the $Z$-factor from the formula~(\ref{GL}) and taking into
account the fact that the spinor $\langle \widehat{P} |$ can be replaced in
both the numerator and denominator of $Z$ by $\langle n | x_{n2}x_{2i}$, we
would obtain $Z^{n,2,i,3,d,2,i,2,n}_{n,2,i;d,2,i,2,n}$. The sequence of
indices $2,i,2$ implies however that one can factor out $x_{2i}^2$.
Since the sequence is present in both the numerator and the denominator,
it can simply be replaced by $2$. Thus we arrive at the form of the $Z$-factor
in the first set of parentheses in~(\ref{HNNMHVa=c=2}).

To verify that equation~(\ref{HNNMHVa=c=2}) is consistent
with~(\ref{H1})
it remains to substitute the $Z$-factors appropriate to the factors
$x_{\widehat{1}d}^2$ and $x_{\widehat{1},i+1}^2$. Doing so we obtain
\begin{equation}
H^{(1)}_{n;2i;2d} =
x_{1i}^2 \Bigl[-Z^{n,d,i,2,n}_{n;i,2,n} \bigl(-Z^{n,2,i,3,d,2,n}_{n,2,i;d,2,n}
P^{2,d-3}_{d,2,n} \bigr) \bigl(-Z^{n,2,i,d+1,d,2,n}_{n,2,i;d,2,n}
P^{d,i-3}_{i,2,n} \bigr)\Bigr] \Bigl[ -Z^{n,i+1,i,2,n}_{n;i,2,n}
P^{i,n-3}_n\Bigr]\,.
\end{equation}
The factor $x_{1i}^2$ gives the required contribution $f_{n;2i}$, while the
factor in the second factor in square brackets is $G^R_{n;2i}$.
The remaining factor in the first set of square brackets is the contribution
from $\widetilde{f}_{n;2i,2d}$ and the other $Z$ and $P$ factors
in~(\ref{H1}).

Now let us look at the terms where $c>2$. We have
\begin{align}
H^{(1)}_{n;2i;cd} =&
x_{1i}^2 \Bigl[x_{\widehat{1} 3}^2 \bigl(-Z^{n,2,i,d,c-1}_{n,2,i;c-1}
P^{2,c-2}_{i,2,n} \bigr) \bigl(-Z^{n,2,i,c+1,d,c,i,2,n}_{n,2,i;d,c,i,2,n}
P^{c,d-3}_{d,c,i,2,n} \bigr) \bigl(-Z^{n,2,i,d+1,d,c,i,2,n}_{n,2,i;d,c,i,2,n}
P^{d,i-3}_{i,2,n} \bigr)\Bigr] \notag\\
&\Bigl[ x_{\widehat{1},i+1}^2 P^{i,n-3}_n\Bigr]\,.
\end{align}
Again, substituting for $x_{\widehat{1}3}^2$ and $x_{\widehat{1},i+1}^2$ we
find agreement with~(\ref{H1}).

Now let us turn our attention to the latter kind of diagrams,
namely the MHV $\times$ NMHV diagrams. From these diagrams we find
\begin{align}
&\sum_{i=4}^{n-3} R^{2}_{n;2i}
\sum_{2\leq c,d < i} \!\!\!
(R^{2i}_{n;cd})^2 H^{(2)}_{n;2i,cd} \notag\\
=& \sum_{i=4}^{n-3} R^{2}_{n;2i}
\sum_{2\leq c,d < i} \!\!\!
(R^{2i}_{n;cd})^2 P^2 G^{\rm NMHV}(\widehat{1},\ldots,-\widehat{P})
G^{\rm MHV}(\widehat{P},\ldots,\overline{n})\,.
\end{align}
As before the sum over $c$ splits into two pieces. For $c=i$ we find
\begin{align}
H^{(2)}_{n;2i;2d} =
x_{1i}^2
\Bigl[ x_{\widehat{1}3}^2 P^{2,i-3}_{i,2,n}\Bigr]
\Bigl[x_{\widehat{1} d}^2 \bigl(-Z^{n,i+1,d,i,n}_{n;d,i,n}
P^{i,d-3}_{d,i,n} \bigr) \bigl(-Z^{n,d+1,d,i,n}_{n;d,i,n}
P^{d,n-3}_{n} \bigr)\Bigr]
\,,
\label{HNNMHV2a=c=2}
\end{align}
while for $c>i$ we find
\begin{align}
H^{(2)}_{n;2i;cd} =
x_{1i}^2
\Bigl[ x_{\widehat{1}3}^2 P^{2,i-3}_{i,2,n}\Bigr]
\Bigl[x_{\widehat{1},i+1}^2 \bigl(-Z^{n,d,c-1}_{n;c-1}
P^{i,c-2}_{n} \bigr) \bigl(-Z^{n,c+1,d,c,n}_{n;d,c,n}
P^{c,d-3}_{d,c,,n} \bigr) \bigl(-Z^{n,d+1,d,c,n}_{n;d,c,n}
P^{d,n-3}_{n} \bigr)\Bigr]
\,.
\end{align}
Making the usual substitutions for the factors of the form $x_{\widehat{1}v}^2$
we find agreement with~(\ref{H2}) in both cases.

To check the terms for $a>2$ we need to consider
$\overline{\rm MHV}_3 \times{\rm NNMHV}_{n-1}$ diagrams. These diagrams
give us
\begin{align}
&\sum_{3 \leq a,b < n} \!\!\!\!
R^{2}_{n;ab}
\Bigl[\sum_{a \leq c,d < b} \!\!\!\!
(R^{ba}_{n;ab;cd})^2 H^{(1)}_{n;ab,cd} + \sum_{b\leq c,d < n} \!\!\!\!
(R_{n;cd}^{ab})^2 H^{(2)}_{n;ab;cd} \Bigr] \notag \\
&= \sum_{3\leq a,b < n} \!\!\!\!
R^{2}_{n;ab}
\Bigl[ \sum_{a\leq c,d < b} \!\!\!\!
(R^{ba}_{n;ab;cd})^2
P^2 H^{(1)}(\widehat{P},\ldots,\overline{n}) + \sum_{b\leq c,d < n} \!\!\!\!
(R^{ab}_{n;cd})^2 H^{(2)}(\widehat{P},\ldots,\overline{n}) \Bigr]\,.
\end{align}
As in the NMHV case, the sum over $a$ splits into a part where $a=3$
and a part where $a>3$.
The calculation is essentially the same as in the NMHV case, with the
factor of $P^2=x_{13}^2$ providing the necessary piece of $f_{n;ab}$ in
both cases. This completes the verification of the formula~(\ref{MNNMHV})
for NNMHV amplitudes.

\section{Discussion of General Tree-Level Amplitudes}

Because of the association between vertices in the rooted
tree diagram~\fig{newYMtree} with individual terms appearing in the
iterative solution of
the recursion relation~(\ref{SBCF}),
it is clear that the procedure applied in the previous section
can be generalized to express an arbitrary
N{}$^p$MHV $n$-graviton super-amplitude
in the form
\begin{equation}
{\mathcal{M}}_n = \sum_{{\mathcal{P}}(2,\ldots,n-1)}
[A^{\rm MHV}(1,\ldots,n)]^2 \sum_{ \{\alpha \} }
[R_\alpha(\lambda_i, \widetilde{\lambda}_i, \eta_i) ]^2
\, G_\alpha(\lambda_i, \widetilde{\lambda}_i)\,,
\end{equation}
where $R_\alpha$ are precisely the same dual superconformal
invariants~(\ref{generalR}) that appear in SYM and $G_\alpha$
are some additional, non-dual conformally invariant,
dressing factors.  Explicit formulas for
the MHV, NMHV, and NNMHV gravity factors are given
respectively in~(\ref{MHVG}), 
(\ref{GNMHV}),
and~(\ref{H1})--(\ref{H2}).

The gravity factors $G_\alpha$ 
for a general amplitude can be worked out on a case-by-case
basis. They always have the form
\begin{equation}
G_{n;a_1b_1;\ldots} = f_{n;a_1b_1} \ldots \, ,
\end{equation}
where $\ldots$ is some combination of $f$, $G^R$ and $G^L$ factors.
The iterative construction of any desired amplitude
is no more difficult than the examples we have already
studied in detail.
Actually one only needs to take care of the factor $f_{n;a_1b_1}$, because
the other parts just go from lower points to higher points
automatically under the
usual rules
\begin{equation}
\langle n|x_{ny} \rightarrow \langle \hat{p}|x_{iy} \rightarrow
\langle n|x_{nj}x_{ji}x_{iy} \label{rule1}\,,
\end{equation}
and
\begin{equation}
\langle n|x_{kl} \rightarrow \langle \hat{p}|x_{kl} \rightarrow
\langle n|x_{nj}x_{ji}x_{kl} \label{rule2}\,,
\end{equation}
as, for example, in going from the NMHV formula~(\ref{GNMHV}) to the
NNMHV formula~(\ref{H1}) and~(\ref{H2}).
The $f$ factors arise at each level for the simple reason that
an extra propagator $P^2$ appears in
on-shell recursion for gravity as compared to the `square' of the
corresponding Yang-Mills result, a fact which we noted
already back in~(\ref{integralexample})
As we already explained
carefully in previous section for the
NMHV case, the factor $f_{n;a_1b_1}$ is needed to satisfy
the recursion relation.


Although it is simple to describe the algorithm for a general
amplitude in words and by appealing to the examples detailed above,
we have not
identified a pattern which would allow us to write
down a general explicit formula, as was done for SYM
in~\cite{Drummond:2008cr}.  
As noted above
each $R_\alpha$ invariant comes with its own $f$-type factor,
and each path 
in~\fig{newYMtree} which ends on a vertex with indices
$a_1b_1;\ldots;a_pb_p$ leads to an associated factor
of the form
\begin{equation}
\label{endfactors}
G^R_{a_1,b_1;\ldots;a_pb_p}
G^L_{a_1b_1;\ldots;a_pb_p}\,,
\end{equation}
where the general $f$, $G^R$ and
$G^L$ are suitably defined following the examples
in the previous section. Specifically we have
\begin{align}
G^L_{n;a_1b_1;...;a_rb_r;ab} & = 
-Z^{n,a_1,b_1,...,a_r,b_r,a+1,b,a,b_r,a_r,...,b_1,a_1,n}_{n,a_1,b_1,...,a_r,b_r;b,a,b_r,a_r,...,b_1,a_1,n} P^{a,b-3}_{b,a,b_r,a_r,...,b_1,a_1,n} \, ,\\
{G}^R_{n;a_1b_1;...;a_rb_r;ab} &=
-Z^{n,a_1,b_1,...,a_r,b_r,b+1,b,a,b_r,a_r,...,b_1,a_1,n}_{n,a_1,b_1,...,a_r,b_r;b,a,b_r,a_r,...,b_1,a_1,n} P^{b,n-3}_{b_r,a_r,...,b_1,a_1,n} \, .
\end{align}
The $f$ factors can be of two types, $\widetilde{f}$ and $\widehat{f}$. The first type are defined as follows,
\begin{align}
\widetilde{f}_{n;a_1b_1;...;a_rb_r;a_rb} &= -Z^{n,a_1,b_1,...,a_r,b_r,b,a_r,b_{r-1},a_{r-1},...,b_1,a_1,n}_{n,a_1,b_1,...,a_{r-1},b_{r-1};b_r,a_r,...,b_1,a_1,n}\,, \\
\widetilde{f}_{n;a_1b_1;...;a_rb_r;ab} &= \bigl(-Z^{n,a_1,b_1,...,b_r,a_r+1,a_r,b_{r-1},a_{r-1},...,b_1,a_1,n}_{n,a_1,b_1,...,a_{r-1},b_{r-1};b_r,a_r,...,b_1,a_1,n}
\bigr)\notag\\
&\phantom{=a\!}\bigl(-Z^{a-1,b,b_r,a_r,...,b_1,a_1,n}_{a-1;b_r,a_r,...,b_1,a_1,n}\bigr) P^{a_r,a-2}_{b_r,a_r,...,b_1,a_1,n} \,\qquad \text{ for } a>a_r .
\end{align}
The second type are given by
\begin{align}
\widehat{f}_{n;a_1b_1;...;a_rb_r;b_rb} &= - Z^{n,a_1,b_1,...,a_{r-1},b_{r-1},b,b_r,a_r,...,b_1,a_1,n}_{n,a_1,b_1,...,a_{r-1},b_{r-1};b_r,a,_r,...,b_1,a_1,n}\,,  \\
\widehat{f}_{n;a_1b_1;...;a_rb_r;ab} &= \bigl(-Z^{n,a_1,b_1,...,a_{r-1},b_{r-1},b_r+1,b_r,a_r,...,b_1,a_1,n}_{n,a_1,b_1,...,a_{r-1},b_{r-1};b_r,a_r,...,b_1,a_1,n} \bigr) \notag \\
&\phantom{=a\!} \bigl(-Z^{n,a_1,b_1,...,a_{r-1},b_{r-1},b,a-1}_{n,a_1,b_1,...,a_{r-1},b_{r-1};a-1} \bigr) P_{b_{r-1},a_{r-1},...,b_1,a_1,n}^{b_r,a-2} \qquad \text{ for } a > b_r\,.
\end{align}

In addition to the factors (\ref{endfactors}), other $G^R$ and $G^L$ factors also appear. If we try
the simplest guess which is that we should be able
to associate these factors to the vertices
in~\fig{newYMtree} such that every
path ending on a given vertex picks up the factors associated
to that vertex, then we find that:
\begin{enumerate}
\item{for each cluster~(see~\fig{cluster}),
the leftmost descendant vertex picks up the same
factors as the parent vertex, and in addition a $G^R$ factor with
the indices of the parent,}
\item{the next descendant vertex to the right is exactly the
same, except that the additional factor is a $G^L$ instead of $G^R$, and
}
\item{going further to the right along the descendant vertices,
there is a more complicated structure
$G^R$ and $G^L$ factors whose indices are modified from those
of the parent.}
\end{enumerate}
We emphasize that we have attempted here only to illustrate
some features of the general structure; in order to determine precisely
the factors which appear for a given path it seems necessary
to work out recursively which kinds of
N{}$^a$MHV$\times$N{}$^b$MHV
factorizations that particular path corresponds to. 

To stress that the algorithm can be simply exploited to generate
higher and higher N${}^p$MHV amplitudes, we give here the formula for
N${}^3$MHV amplitudes:
\begin{align}
&M^{{\rm N}^3{\rm MHV}}(1,\ldots,n) =
[A^{\rm MHV}(1,\ldots,n)]^2
\sum_{2 \leq a_1, b_1 < n} \!\!\!\!\!\! R^2_{n;a_1b_1} \Bigl[  \notag \\
&\sum_{a_1 \leq a_2,b_2 < b_1} \!\!\!\!\!\! (R_{n;a_1b_1;a_2b_2}^{b_1a_1})^2
\Bigl(\sum_{a_2 \leq a_3,b_3 < b_2} \!\!\!\!\!\!
(R_{n;a_1b_1;a_2b_2;a_3b_3}^{a_1b_1b_2a_2})^2
G^{(1)}_{n;a_1b_1;a_2b_2;a_3b_3} \nonumber \\
& \qquad\qquad +
\sum_{b_2 \leq a_3,b_3 < b_1} \!\!\!\!\!\!
(R_{n;a_1b_1;a_3b_3}^{a_1b_1a_2b_2})^2 
G^{(2)}_{n;a_1b_1;a_2b_2;a_3b_3}+
\sum_{b_1 \leq a_3,b_3 < n} \!\!\!\!\!\!
(R_{n;a_3,b_3}^{a_1b_1})^2
G^{(3)}_{n;a_1b_1;a_2b_2;a_3b_3} \Bigr) \nonumber \\
&+ \sum_{b_1 \leq a_2,b_2 < n} \!\!\!\!\!\!
(R_{n;a_2,b_2}^{a_1b_1})^2 \Bigl(\sum_{a_2 \leq a_3,b_3 < b_2} \!\!\!\!\!\!
(R_{n;a_2b_2;a_3b_3}^{b_2a_2})^2
G^{(4)}_{n;a_1b_1;a_2b_2;a_3b_3}
+ \sum_{b_2 \leq a_3,b_3 < n} \!\!\!\!\!\!
(R_{n;a_3,b_3}^{a_2b_2})^2
G^{(5)}_{n;a_1b_1;a_2b_2;a_3b_3}\Bigr)
\Bigr]\,.
\end{align}
The five different $G$-factors are in correspondence with the
five different vertical paths from the root
node to the vertices on the lowest row explicitly shown in~\fig{newYMtree}.
Explicitly they are given by
\begin{align}
G^{(1)}_{n;a_1b_1;a_2b_2;a_3b_3}&=
f_{n;a_1b_1}
\widetilde{f}_{n;a_1b_1;a_2b_2}
\widetilde{f}_{n;a_1b_1;a_2b_2;a_3b_3}
G^R_{n;a_1b_1}
G^R_{n;a_1b_1;a_2b_2}
G_{n;a_1b_1;a_2b_2;a_3b_3}\,,
\\
G^{(2)}_{n;a_1b_1;a_2b_2;a_3b_3}&=
f_{n;a_1b_1}
\widetilde{f}_{n;a_1b_1;a_2b_2}
\widetilde{{f}}_{n;a_1b_1;a_2b_2;a_3b_3}
G^R_{n;a_1b_1}
G^L_{n;a_1b_1;a_2b_2}
G_{n;a_1b_1;a_3b_3},\
\\
G^{(3)}_{n;a_1b_1;a_2b_2;a_3b_3}&=
f_{n;a_1b_1}
\widetilde{f}_{n;a_1b_1;a_2b_2}
\widehat{f}_{n;a_1b_1;a_3b_3}
G_{n;a_1b_1;a_2b_2}
G_{n;a_3b_3}\,,
\\
G^{(4)}_{n;a_1b_1;a_2b_2;a_3b_3}&=
f_{n;a_1b_1}
\widehat{f}_{n;a_1b_1;a_2b_2}
\widehat{{f}}_{n;a_1b_1;a_2b_2;a_3b_3}
G^L_{n;a_1b_1}
G^R_{n;a_2b_2}
G_{n;a_2b_2;a_3b_3},\
\\
G^{(5)}_{n;a_1b_1;a_2b_2;a_3b_3}&=
f_{n;a_1b_1}
\widehat{f}_{n;a_1b_1;a_2b_2}
\widehat{f}_{n;a_1b_1;a_2b_2;a_3b_3}
G^L_{n;a_1b_1}
G^L_{n;a_2b_2}
G_{n;a_3b_3}\,,
\end{align}
where $G$ is shorthand for
$G^L\times G^R$ (with the same subscripts on both).

The expressions we have found can certainly be used in the calculation of loop amplitudes in supergravity. It is straightforward to apply the generalized unitarity technique in a manifestly supersymmetric way \cite{Drummond:2008bq,Brandhuber:2008pf,ArkaniHamed:2008gz}; the basic ingredients in this procedure are the tree-level super-amplitudes. 

It would of course be extremely interesting
to unlock the general pattern
of $G$-factors to allow one to write down a general explicit formula.
It would also be interesting to
see if the SUGRA `bonus relations'~\cite{ArkaniHamed:2008gz,Spradlin:2008bu}
could be usefully exploited beyond MHV amplitudes.
There is no doubt that much additional structure
remains to be found.
Hopefully, much simpler and more beautiful formulas await than the
ones obtained here.
Certainly
this should be the case if the notion that SUGRA amplitudes are even simpler
than those of SYM is to come to full fruition.

\section*{Acknowledgments}

We are grateful to
Nima Arkani-Hamed, Zvi Bern, Lance Dixon,
Henriette Elvang, Dan Freedman, Johannes Henn, Chrysostomos Kalousios,
Steve Naculich and Cristian Vergu for stimulating
discussions and helpful correspondence.
This work was supported in part by the
French Agence Nationale de la Recherche under
grant ANR-06-BLAN-0142 (JMD), the
US Department of Energy under contract
DE-FG02-91ER40688 (MS (OJI) and AV), and the
US National Science Foundation under grants PHY-0638520 (MS) and
PHY-0643150
CAREER PECASE (AV).

\appendix

\section{Conventions}\label{conventions}

Here we give some formulae to establish the conventions we are using for the
two-component spinors.
We have
\begin{equation}
x_{\alpha \dot \alpha} \equiv x_{\dot{\alpha} \alpha}
=(\sigma^\mu)_{\alpha \dot \alpha} x_\mu, \hspace{30pt} x^{\dot \alpha \alpha}
\equiv x^{\alpha \dot\alpha}=
(\widetilde{\sigma}^\mu)^{\dot \alpha \alpha}x_{\mu}\,,
\end{equation}
\begin{equation}
x^2 = x^\mu x_\mu
= \tfrac{1}{2} x_{\alpha \dot\alpha}x^{\dot\alpha \alpha},
\hspace{30pt}
x_{\alpha \dot \alpha}x^{\dot \alpha \beta} = \delta_\alpha^\beta x^2,
\hspace{30pt}
x^{\dot \alpha \alpha} x_{\alpha \dot\beta} =
\delta^{\dot \alpha}_{\dot \beta} x^2\,.
\end{equation}
For the commuting spinors, the following notation has been used,
\begin{equation}
p_i^{\alpha \dot \alpha} =
\lambda_i^{\alpha} \widetilde{\lambda}_i^{\dot \alpha},
\hspace{20pt}
\lambda_{i \alpha} = \lambda_i^{\beta} \epsilon_{\beta \alpha},
\hspace{20pt}
\lambda_i^\alpha = \epsilon^{\alpha \beta} \lambda_{i \beta},
\hspace{20pt}
\widetilde{\lambda}_{i \dot \alpha} =
\widetilde{\lambda}_i^{\dot \beta} \epsilon_{\dot{\beta} \dot{\alpha}},
\hspace{20pt}
\widetilde{\lambda}_i^{\dot \alpha} =
\epsilon^{\dot{\alpha} \dot{\beta}} \widetilde{\lambda}_{i \dot \beta}\,,
\end{equation}
where $\epsilon^{\alpha \beta}$ and $\epsilon^{\dot{\alpha} \dot{\beta}}$
are antisymmetric tensors.
For the contractions of these spinor variables we write for example
\begin{align}
\ket{i}{j} = \lambda_i^{\alpha} \lambda_{j \alpha},
\hspace{30pt}
\bra{i}{j} = \widetilde{\lambda}_i^{\dot \alpha}
\widetilde{\lambda}_{j \dot\alpha}\,,
\end{align}
\begin{equation}
\langle i | x | j] =
\lambda_i^{\alpha} x_{\alpha \dot\alpha} \widetilde{\lambda}_j^{\dot \alpha},
\hspace{30pt}
\langle i | x_1 x_2 \ldots x_{2 m} | j\rangle
= \lambda_i^{\alpha} x_{1 \alpha \dot \alpha} x_2^{\dot \alpha \beta}
\ldots x_{2 m}^{\dot \gamma \gamma} \lambda_{j \gamma}\,.
\end{equation}
For the dual coordinates we use
\begin{align}
p_i^{\alpha \dot \alpha} 
&=x_i^{\alpha \dot\alpha} - x_{i+1}^{\alpha \dot\alpha},
\hspace{30pt} x_{n+1} \equiv x_1\,. 
\end{align}
Similarly for the Grassmann odd dual coordinates we have
\begin{align}
q_i^{\alpha A} = \lambda_i^{\alpha} \eta_i^A = \theta_i^{\alpha A} - \theta_{i+1}^{\alpha A}.
\end{align}

\section{NMHV Graviton Amplitudes}

Here we provide some additional details regarding the
formula for 
general NMHV super-amplitudes proven in section III.B, 
\begin{equation}
{\mathcal{M}}_n^{\rm NMHV}
= \sum_{{\mathcal{P}}(2,\ldots,n-1)}
\left[ \frac{ \delta^{(8)}(q) } { \ket{1}{2} \cdots \ket{n}{1} }
\right]^2
\sum_{s=2}^{n-3} \sum_{t=s+2}^{n-1} R_{n;st}^2 G^{\rm NMHV}_{n;st}\,,
\end{equation}
where the $G$-factor is given in~(\ref{GNMHV})
and the dual superconformal invariant is
\begin{equation}
R_{n;st} = \frac{ \ket{s}{s-1} \ket{t}{t-1}
\delta^{(4)}(\Xi_{n;st})}
{x_{st}^2 \langle n|x_{ns} x_{st} |t \rangle
\langle n| x_{ns} x_{st} |t-1\rangle
\langle n| x_{nt} x_{ts} |s\rangle
\langle n|x_{nt} x_{ts} |s-1\rangle}
\end{equation}
in terms of~\cite{Drummond:2008vq,Drummond:2008cr}
\begin{equation}
\Xi_{n;st} = \langle n| \left[
x_{ns} x_{st} \sum_{i=t}^{n-1} |i\rangle \eta_i
+ x_{nt} x_{ts} \sum_{i=s}^{n-1} |i\rangle \eta_i
\right]\,.
\end{equation}

In order to extract the $n$-particle NMHV graviton amplitude
from this superspace expression we should perform the integral
over $d^8 \eta_i$ for the three negative helicity gravitons $i$.
It is convenient to choose particles 1 and $n$ to be two of these
three since $\Xi_{n;st}$ does not depend on $\eta_1$ or $\eta_n$.
These two variables appear only inside the supermomentum conserving
delta function $\delta^{(16)}(q)$ which may be put into the
form~\cite{Drummond:2008bq,Drummond:2008cr}
\begin{equation}
\delta^{(16)}(q) = \ket{1}{n}^8
\,
\delta^{(8)}
\left( \eta_1^A + \sum_{i=2}^{n-1} \frac{ \ket{n}{i} }{\ket{n}{1} }
\eta_i^A \right)
\delta^{(8)}
\left( \eta_n^A + \sum_{i=2}^{n-1} \frac{ \ket{i}{1} }{ \ket{n}{1} }
\eta_i^A \right)
\,.
\end{equation}
The $d^8\eta_1 d^8\eta_n$ integrals are then trivial, leading to
\begin{equation}
\label{NMHVgraviton}
{\mathcal{M}}(1^-,2^-,3^+,\ldots,n^-)
= \int d^8 \eta_2 \sum_{{\mathcal{P}}(2,\ldots,n-1)}
\sum_{s=2}^{n-3}
 \sum_{t=s+2}^{n-1}
\left[
\frac{  \ket{1}{n}^4 \, R_{n;st}}{\ket{1}{2} \cdots \ket{n}{1}}\right]^2
G_{n;st}^{\rm NMHV}\,.
\end{equation}
Here we have chosen, without loss of generality,
particle 2 to be the third negative helicity graviton.

The analogous NMHV gluon amplitude simplifies further due to the
fact that $R_{n;st}$ only depends on $\eta_2$ when $s=2$; thus
performing the integral
eliminates the sum over $s$~\cite{Drummond:2008cr}.
Here in the case of gravity
it is unfortunately cumbersome to proceed analytically because
the sum over
permutations in~(\ref{NMHVgraviton}) generates
many terms, even for
the simplest nontrivial case $n=6$ where the sum over $s$ and $t$
produces just three terms and the
corresponding $G$-factors simplify considerably,
\begin{align}
G^{\rm NMHV}_{6;24} &=
+ (p_1 + p_2 + p_3)^2
\bra{1}{2} \ket{2}{3} \bra{4}{5} \ket{5}{6}
\frac{\langle 6|5+4|3] \, \langle 4|3+2|1]}
{\langle 6|5+4|1] \, \langle 6|3+2|1]}\,
\\
G^{\rm NMHV}_{6;25} &=
+ (p_5 + p_6)^2
\ket{2}{3} \bra{3}{4}
\frac{ \langle 6|1+2|3+4|5+6| 1] }
{ \ket{5}{6} \bra{1}{5} }
\frac{ \langle 4|5+6|1 ] }{\langle 2|5+6| 1]}\,,
\\
G^{\rm NMHV}_{6;35} &=
- (p_1+p_2)^2 \ket{3}{4} \bra{4}{5} \ket{5}{6}^2
\frac{ \langle 2|3+4|5] \,
\langle 6|1+2|3 ]
}
{\ket{2}{6}\, \langle 6|1+2|3+4|6 \rangle}\,.
\end{align}
Therefore we do not provide explicit analytic formulas for
graviton amplitudes, which instead may be evaluated numerically as needed.
We have checked numerically that our expression agrees with
other representations for the $n=6$ particle NMHV graviton
amplitude in the literature (see for
example~\cite{Cachazo:2005ca,Bianchi:2008pu}).

\end{document}